41

# Demonstration of surface-engineered oxidation-resistant Nb–Nb thermocompression bonding toward scalable superconducting quantum computing architectures

Harsh Mishra,[1] Yusuke Kozuka,[2,3] Sathish Bonam,[4] Jun Uzuhashi,[5] Praveenkumar Suggisetti,[7] Tadakatsu Ohkubo,[6] and Shiv Govind Singh[1*]

[1] *Department of Electrical Engineering, Indian Institute of Technology Hyderabad, Sangareddy, Telangana, 502285, India*
[2] *Research Center for Materials Nanoarchitectonics, National Institute for Materials Science (NIMS), Tsukuba 305-0044, Japan*
[3] *WPI Advanced Institute for Materials Research, Tohoku University, 2-1-1 Katahira, Aoba-ku, Sendai 980-8577, Japan*
[4] *Tyndall National Institute, Lee Maltings Complex Dyke Parade, Cork, T12 R5CP, Ireland*
[5] *Research Network and Facility Services Division, National Institute for Materials Science (NIMS), Tsukuba 305-0047, Japan*
[6] *Research Center for Magnetic and Spintronic Materials, National Institute for Materials Science (NIMS), Tsukuba 305-0047, Japan*
[7] *Bhabha Atomic Research Centre, Mumbai, India*

Scalable quantum computing currently requires a large array of qubit integration, but present two-dimensional interconnects face challenges such as wiring congestion, electromagnetic interference, and limited cryogenic space. To overcome this challenge, implementing three-dimensional (3D) vertical architectures becomes crucial. Niobium (Nb), due to its excellent superconducting characteristics and strong fabrication process compatibility, stands out as a prime material choice. The main challenge in Nb–Nb bonding is the presence of an oxide layer at the interface, even after post-bonding annealing across various bonding methods. The native Nb oxide forms rapidly in air, creating a resistive barrier to supercurrent flow and introducing two-level system losses that degrade qubit coherence while increasing the overall thermal budget. These issues show the need for effective surface engineering to suppress oxidation during bonding. This study introduces an ultrathin gold (Au) capping layer as a passivation strategy to prevent oxygen incorporation at the Nb surface. This approach enables low-temperature Nb–Nb thermocompression bonding at 350 °C under a reduced bonding pressure of 0.495 MPa. Detailed microstructural and interfacial analyses confirm that Au passivation effectively suppresses oxide formation and hence enhances bonding uniformity and strength with keeping the superconductivity, establishing a robust route toward low-temperature, low-pressure Nb–Nb bonding for scalable 3D superconducting quantum computing architectures.

**Keywords**: Oxidation Resistant, Thermocompression Bonding, Niobium (Nb), Superconductivity, Quantum Computing

Quantum computing is still in the initial phase of its engineering evolution. Although the fundamental physics is well established, translating these principles into practical, computational devices presents notable technological hurdles. At present, experimental systems are capable of controlling only a relatively small number of qubits,[1] pointing to the demanding engineering efforts necessary for a practical, large-scale quantum computer.

Several technological developments are still required. Currently, three-dimensional (3D) integration and packaging of superconducting quantum circuits is becoming a key research focus for enabling larger and more densely integrated qubit architectures.[2] These architectures rely on superconducting through-silicon vias (TSVs) and interposers, which minimize resistive losses and heat dissipation at cryogenic temperatures, unlike conventional copper interconnects. Most processors operate below the critical temperature ($T_c$) of superconducting materials, where zero resistance is achieved. While copper integration is well studied,[3-7] scalable integration with the superconducting materials remains a key challenge. Several studies explore multiple materials as an integration, such as

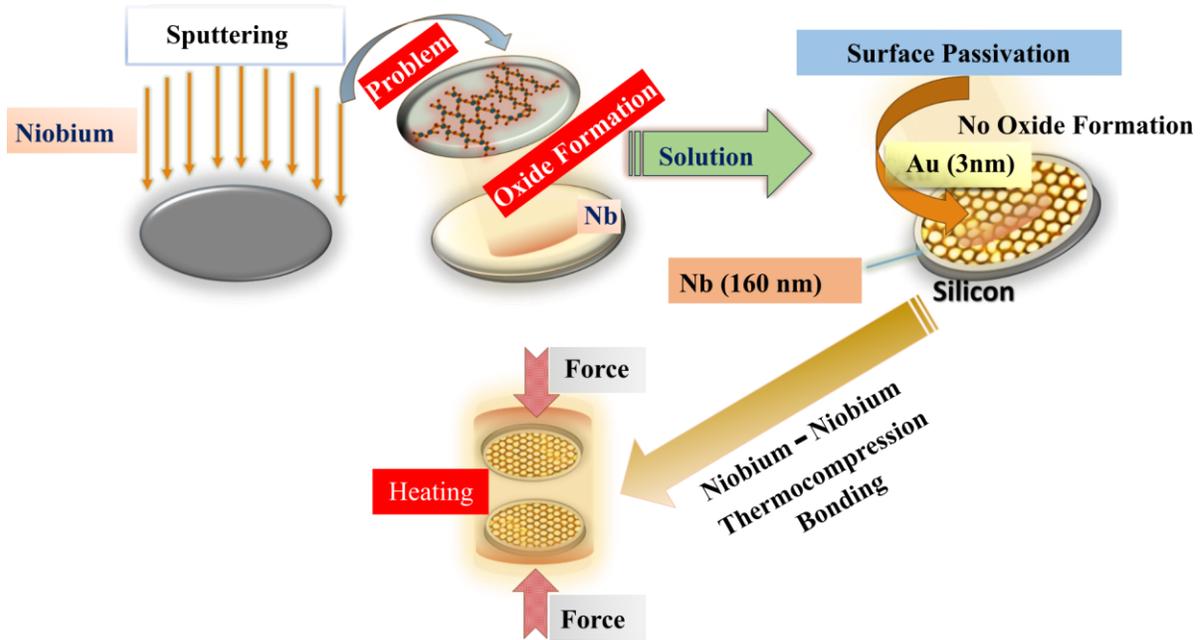

Figure 1. Schematic of Au-passivated Nb-Nb bonded structure and the process of thermocompression bonding.

Ta, Nb, NbN, and Indium.[8-12] Recently, P. Renaud *et al.* presented the hydrophilic direct Nb-Nb bonding driven by chemical mechanical polishing (CMP) at ambient temperature under atmospheric pressure, followed by two hours of annealing at 400 °C to facilitate oxide dissolution. Post-bonding annealing can result in improved electrical conduction and mechanical strength, while also facilitating the creation of metallic bonds between the upper and lower metals. However, there is an enrichment of oxygen observed at the bonding interface even after annealing.[13] The oxide layer leads to higher resistance and causes signal loss in the microwave frequencies used in superconducting quantum computers. A method is required to effectively reduce or eliminate oxide formation at the Nb–Nb bonding interface.

To address this issue, this study presents an ultrathin gold (Au) capping layer as an effective method for protecting Nb films from oxidation while improving bonding quality using the thermocompression technique. This passivation approach helps in low-temperature Nb-Nb thermocompression bonding at 350°C under a significantly reduced bonding pressure of 0.495 MPa, which results in lowering processing constraints while preserving superconducting integrity. A comprehensive materials and interface characterization is conducted using multiple analytical techniques to assess the impact of passivation on bonding quality, interfacial uniformity, and oxide formation suppression. The results demonstrate that ultrathin inert metal passivation not only enhances Nb-Nb bond integrity but also enables more controlled and repeatable bonding at reduced thermal budgets. These findings establish a viable route for low-temperature and low-pressure Nb-Nb thermocompression bonding, paving the way for the development of 3D-integrated superconducting chips and scalable quantum computing architectures with improved coherence performance and fabrication efficiency.

Thin films were deposited at room temperature using a DC sputtering system (AJA International Inc., USA) for the thermocompression bonding process. The deposition was carried out under an Ar pressure of 3 mTorr, with sputtering powers of 150 W for Nb and 75 W for Au. Sequential deposition was carried out on the pre-treated silicon wafers, starting with a 160 nm Nb layer, followed by an ultrathin Au passivation layer, all without breaking vacuum. Thin film samples for X-ray diffraction (XRD) and superconductivity measurement were prepared using a high resistivity silicon wafer, where the thickness of 100 nm was optimized for Nb with 5 mTorr Ar pressure at a 200 W sputtering power, similarly, Au of 3 nm at the same Ar pressure at 20 W. Throughout the paper, the Au-passivated samples are denoted as Nb/Au. The thicknesses of the Au passivation layer ranges from 1 to 7 nm.

Atomic force microscopy (AFM, Bruker Icon ScanAsyst system) was employed to access the surface morphology. Crystal structures of the thin films were characterized by XRD (SmartLab II, Rigaku Co.) equipped with a Cu $K\alpha$ radiation source ($\lambda$ = 1.5418 Å). Superconducting properties were measured using a cryostat equipped with a 14 T superconducting magnet (PPMS Dynacool™, Quantum Design).

Contact angle measurements were performed using the video contact angle (VCA) Optima setup (provided by AST Products, Inc.) for evaluating surface wettability over time due to gradual surface oxidation. Thermocompression bonding was carried out using a commercially available wafer bonder (Applied Microengineering Limited, AML-AWB-04), capable of handling wafers up to 4 inches. The bonding process was performed at 300 °C, 350 °C, and 400 °C under a pressure of 0.495 MPa for 120 minutes.

We first analyze the surface roughness of Au-passivated Nb samples with varying the thicknesses of the Au passivation layer using AFM as shown in Fig. 2. To establish a reference for comparison, the surface roughness of an unpassivated Nb film was measured, resulting in a root-mean-square (RMS) roughness ($R_q$) of 1.8 nm. This value serves as a baseline for assessing the impact of Au passivation. As summarized in Fig. 2(b), the RMS roughness varies with different Au layer thicknesses. Notably, the RMS roughness is minimum around an Au thickness of 3 nm, which is about 0.75 nm. As shown in Fig. 2(a), the Nb surface passivated with a 3 nm Au layer exhibits a smooth surface. This reduction demonstrates the role of Au passivation to prevent the roughening of Nb surface, which enhances bonding quality with lower pressures as discussed later. A distinct trend emerges at greater passivation thicknesses, where the RMS roughness increases to 1.5 nm in the case of 7 nm-thick Au layer, as shown in Fig. 2(b). This may be due to the coalescence nature of Au once the film becomes thicker.

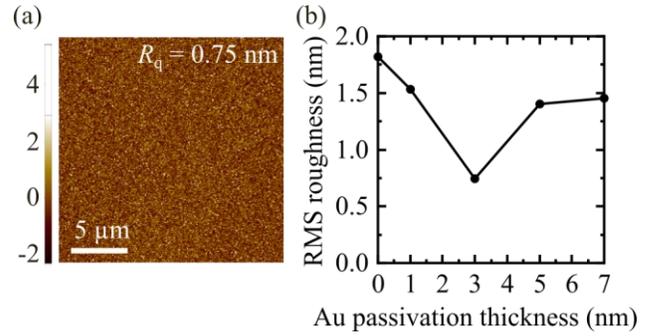

Figure 2. (a) An atomic force microscopy image of Nb passivated with 3 nm-thick Au. (b) Root mean square (RMS) roughness as a function of passivation layer (Au) thickness.

To check the effect of the Au passivation, contact angle is measured to obtain an insight into surface oxidation, which impacts bonding reliability. As shown in Fig. S1 in the supplementary material, high contact angle values of the sample passivated with 3 nm-thick Au remain practically unchanged with aging closer to 120 hours, while the bare Nb thin film with its native oxide exhibits a change in contact angle of about 41.6° to 83.4° over the duration of measurement. Such a contact angle indicates a rather hydrophilic surface, which can be expected for a thin, clean residual oxide layer that is most likely present on the Nb film. Similar results are reported with Nb films passivated with self-assembled monolayer.[14] Hence, this results further supports that Au passivation is effective to suppress the oxide formation.

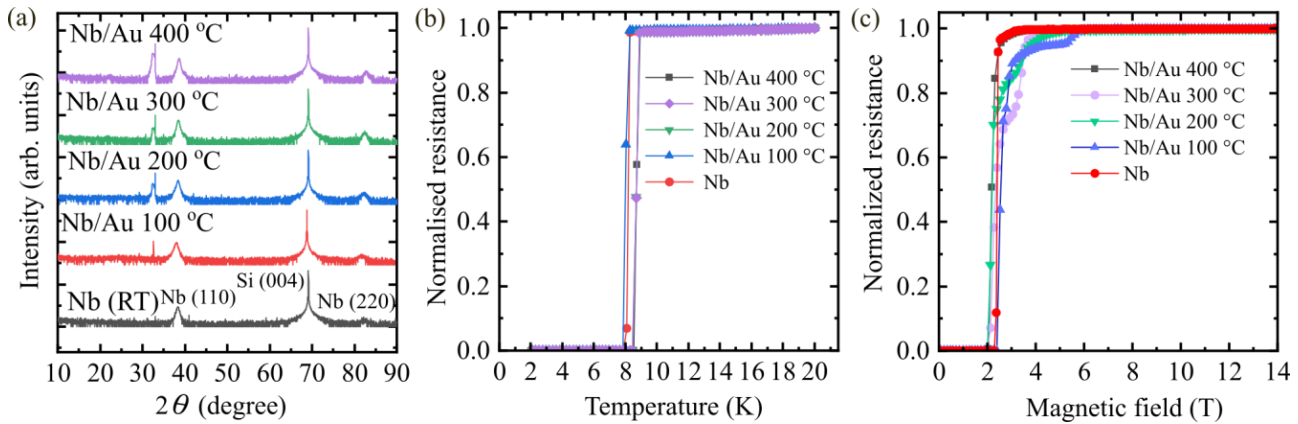

Figure 3. (a) XRD patterns of Nb and Nb/Au thin films. (b) Temperature and (c) magnetic field dependences of electrical resistance normalized by their normal resistances.

To characterize the crystal structures of Nb thin films passivated with Au, XRD is measured as shown in Fig. 3(a). The XRD patterns confirm that all samples exhibit a body-centered cubic (bcc) structure of niobium, with a dominant Nb (110) out-of-plane orientation evidenced by the peak near $2\theta \approx 38°$.[15-17] Based on XRD, Nb thin films passivated with Au show the same crystal structure as one without Au passivation, indicating that the structure of Nb is almost intact to thin Au deposition. The peak around $2\theta \approx 33°$ corresponds to the forbidden Si (002) diffraction and may result from the slight modification of the Si surface by Nb deposition.

We then evaluate the superconducting characteristics as shown in Figs. 3(b) and 3(c). The Nb film without passivation exhibits the transition temperature of ~8.1 K, slightly below the bulk value of ~ 9.26 K.[18] This decrease likely arises from several contributing factors, such as oxygen incorporation during film deposition, strain imposed by the substrate, and surface oxidation that occurs when the film is exposed to air. Similar behaviour has been observed in other reports where Nb films grown by high power impulse magnetron sputtering (HiPIMS) with thicknesses of 40 nm and 155 nm.[19] They observed that the $T_c$ varied between approximately 7 and 9 K depending on the substrate type and film thickness, showing how growth conditions and substrate choice affect superconducting behaviour. For the Nb/Au film annealed at 100 °C, the $T_c$ is similar to the bare Nb film, whereas for Nb/Au films annealed at higher temperature, the $T_c$ improves to ~8.7 K as shown in Fig. 3(b). Interestingly, the bare Nb sample exhibits the critical magnetic field of around 2.36 T, while the Nb/Au film annealed at 100 °C exhibits 2.53 T; similarly, the Nb/Au samples annealed at higher temperatures exhibit ~ 2.2 T, as shown in Fig. 3(c). The critical magnetic field in the Nb/Au film annealed at 100 °C is enhanced compared to bare Nb, despite suppressed $T_c$. Although it may be counterintuitive, it is relevant to dirty limit superconductors, where electron scattering reduces the Ginzburg-Landau coherence length $\xi_{GL}(0)$:[20-21]

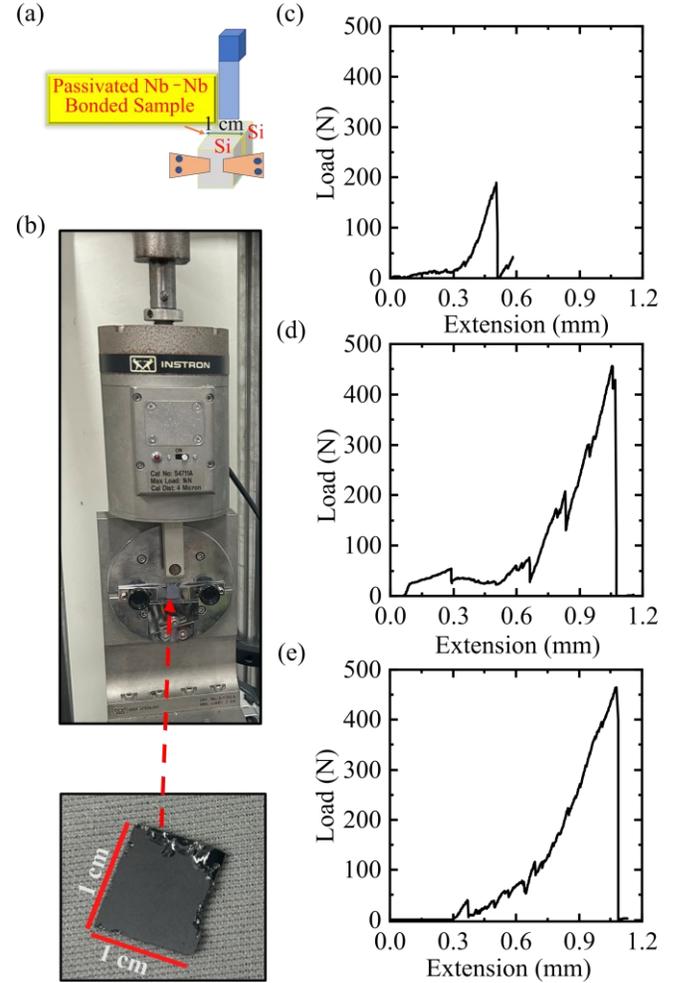

Figure 4. (a) Schematic illustration of the shear bond strength measurement setup. (b) Experimental setup showing post-test interface inspection and bond strength measurements of samples bonded at (c) 300 °C, (d) 350 °C, and (e) 400 °C.

$$\mu_o H_{c2}(0) = \frac{\Phi_0}{2\pi\xi_{GL}^2(0)}, \quad (1)$$

where $\mu_o$ is the vacuum permeability, $H_{c2}(0)$ is the zero-temperature upper critical field, and $\Phi_0 = h/2e$ is the magnetic flux quantum. The shorter the coherence length, the greater the enhanced scattering, possibly arising from interfacial disorder, resulting in a relatively high critical magnetic field. Similarly, at higher annealing temperature, lattice defects may be partially removed to increase the mean free path ($\ell$), and therefore, the effective Ginzburg-Landau coherence length according to:[22-24]

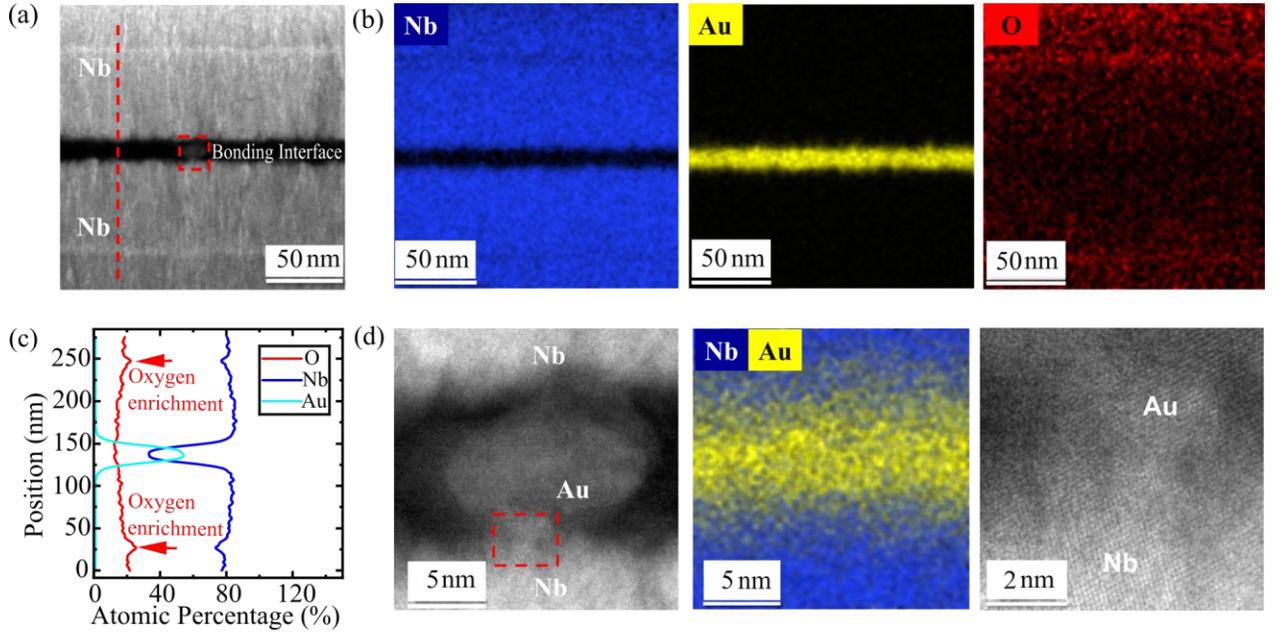

Figure 5. (a) BF-STEM image of the Au passivated Nb-Nb bonding interface bonded at 350 °C and (b) corresponding EDS maps revealing the absence of oxygen at the bonding interface. (c) EDS line profile across the bonding interface along the dash line marked in (a). (d) (Left) High-magnification BF-STEM image at the marked square region in (a) at the interface. (Middle) The corresponding EDS maps verify the elemental distribution of Nb and Au across the interface. (Right) The high-resolution BF-STEM image from the square marked region in the BF-STEM image confirms the structural integrity of the bonded layers.

$$\xi_{GL} \propto \sqrt{\xi_0 \ell}, \qquad (2)$$

where $\xi_0$ is the BCS coherence length related to the transition temperature. Hence, the critical magnetic field decreases as increasing $\xi_{GL}$ in Nb/Au samples annealed above 200 °C.

Having characterized the basic properties of Nb films with and without Au passivation, we have conducted thermocompression bonding at several temperatures ranging from 300 to 400 °C, and characterized the bonding strength, as shown in Fig. 4(a). Bond strength is crucial for semiconductor applications to ensure that the bonded interface can withstand mechanical stress during wafer dicing and back thinning. Test samples (~1 cm × 1 cm), as shown in Fig. 4(b), are manually diced from a 2-inch bonded wafer for analysis. Shear tests recorded maximum forces of 189 N (~65.39 MPa) at 300 °C, 456 N (~157.7 MPa) at 350 °C, and 464 N (~160.5 MPa) at 400 °C, as shown in Figs. 4(c), 4(d), and 4(e), respectively, demonstrating consistent results across multiple specimens. From the bond strength analysis, we observe a significant increase in bond strength from 300 °C to 350 °C, while a comparable bond strength is observed at 400 °C compared to 350 °C. Therefore, a minimum bonding temperature of 350 °C is required to achieve strong superconducting joint adhesion, consistent with the bond strength reported in previous studies.[3-5]

To gain deeper insight into the grain growth across the bonded interface, cross-sectional bright-field scanning transmission electron microscopy (BF-STEM) combined with energy-dispersive X-ray spectroscopy (EDS) is conducted on the sample bonded at 350 °C (Spectra Ultra STEM at 300 kV, *Thermo Fisher Scientific*). The cross-sectional thin foil specimen of 25-nm-thickness was prepared by FIB-SEM system based on back-scattered electron intensity method (Helios5UX, *Thermo Fisher Scientific*).[25] The cross-sectional BF-STEM image of the Au-passivated Nb–Nb bonded sample is shown in Fig. 5(a), while the corresponding EDS elemental maps in Fig. 5(b) confirm the presence of Nb and Au with no detectable oxygen at the bonding interface. It should be noted that some amount of O is unavoidably detected as background by EDS due to the natural oxidization on both surfaces of thin specimen during specimen transfer from FIB-SEM to TEM system. A slight oxygen enrichment observed in line EDS away from the interface, as shown in Fig. 5(c), is attributed to an intermittent pause during Nb layer deposition. This observation confirms that the bonding interface itself is oxygen-free. To further examine the void morphology at the interface, a high-magnification BF-STEM image consisting of an EDS map of the same region is shown in Fig. 5(c), which was obtained at the square region marked in Fig. 5(a). The image highlights the formation of a large Au crystal grain across the interface (marked by a square), indicating a void-free interface, which was further confirmed through a high-resolution BF-STEM image at the same square marked in high-magnification BF-STEM image. This contributes to the observed high bond strength. The EDS map shown in Fig. 5(d) corroborates that Nb and Au interdiffuse by ~15 nm at the interface, possibly originating from thermal diffusion during the thermocompression bonding at 350 °C. However, this extent of diffusion is negligible compared to the total thickness of Nb and superconducting coherence length of Nb, and the superconducting properties are nearly intact.

This work demonstrates a notable advancement in Nb–Nb thermocompression bonding, using an optimized 3 nm ultrathin Au passivation layer, which results in robust integration at 350 °C. The passivation effectively suppresses Nb oxidation, confirmed by TEM–EDS, and reduces surface roughness to 0.75 nm without the chemical mechanical polishing process. Post-bonding BF-STEM imaging shows a clean, seamless interface without oxide and voids, while shear tests reveal high bond strengths of 157.7 MPa at 350 °C and 160.5 MPa at 400 °C, compared to only 65.39 MPa at 300 °C. These results identify 350 °C as the minimum temperature needed to achieve reliable, high-quality bonding. Overall, ultrathin Au passivation significantly enhances bond performance and supports scalable, low-temperature multilayer integration for quantum computing applications.

See the supplementary material for the details related to contact angle measurement.

The author, H.M., would like to express appreciation for the financial support from the JST Lotus Fellowship, TCS Research Fellowship, and the technical support of the Nano-X fabrication Lab. Y.K. acknowledges the support by JST FOREST Program (Grant Number: JPMJFR203D). A part of this work was supported by the Electron Microscopy Unit and the Surface and Bulk Analysis Unit, National Institute for Materials Science (NIMS), and "Advanced Research Infrastructure for Materials and Nanotechnology in Japan (ARIM)" of the Ministry of Education, Culture, Sports, Science and Technology (MEXT) (Proposal Number JPMXP1225NM5078). MANA and AIMR are supported by World Premier International Research Center Initiative (WPI), MEXT, Japan.

## AUTHOR DECLARATIONS

### CONFLICT OF INTEREST

The authors have no conflicts to disclose.

### AUTHOR CONTRIBUTIONS

**Harsh Mishra:** Investigation (lead); Data curation (equal); Formal analysis (equal); Methodology (equal); Validation (equal); Visualization (equal); Writing - original draft (lead); Writing - review & editing (equal). **Yusuke Kozuka:** Conceptualization (equal); Methodology (equal); Writing - review & editing (equal); Resources (equal); Project administration (equal); Funding acquisition (equal); Supervision (equal). **Sathish Bonam:** Investigation (equal); Conceptualization (equal); Methodology (equal); Writing - review & editing (equal). **Jun Uzuhashi:** Methodology (equal); Writing - review & editing (equal); Resources (equal). **Praveenkumar Suggisetti:** Conceptualization (equal); Methodology (equal); Writing - review & editing (equal). **Tadakatsu Ohkubo:** Methodology (equal); Writing - review & editing (equal); Resources (equal). **Shiv Govind Singh:** Conceptualization (lead); Methodology (equal); Writing - review & editing (equal); Resources (equal); Project administration (equal); Funding acquisition (equal); Supervision (lead).

### DATA AVAILABILITY STATEMENT

The data that support the findings of this study are available from the corresponding authors upon reasonable request.

# Demonstration of surface-engineered oxidation-resistant Nb–Nb thermocompression bonding toward scalable superconducting quantum computing architectures (Supplementary material)

Harsh Mishra,[1] Yusuke Kozuka,[2,3] Sathish Bonam,[4] Jun Uzuhashi,[5] PraveenKumar Suggisetti,[7] Tadakatsu Ohkubo,[6] and Shiv Govind Singh[1*]

[1] Department of Electrical Engineering, Indian Institute of Technology Hyderabad, Sangareddy, Telangana, 502285, India
[2] Research Center for Materials Nanoarchitectonics, National Institute for Materials Science (NIMS), Tsukuba 305-0044, Japan
[3] WPI Advanced Institute for Materials Research, Tohoku University, 2-1-1 Katahira, Aoba-ku, Sendai 980-8577, Japan
[4] Tyndall National Institute, Lee Maltings Complex Dyke Parade, Cork, T12 R5CP, Ireland
[5] Research Network and Facility Services Division, National Institute for Materials Science (NIMS), Tsukuba 305-0047, Japan
[6] Research Center for Magnetic and Spintronic Materials, National Institute for Materials Science (NIMS), Tsukuba 305-0047, Japan
[7] Bhabha Atomic Research Centre, Mumbai, India

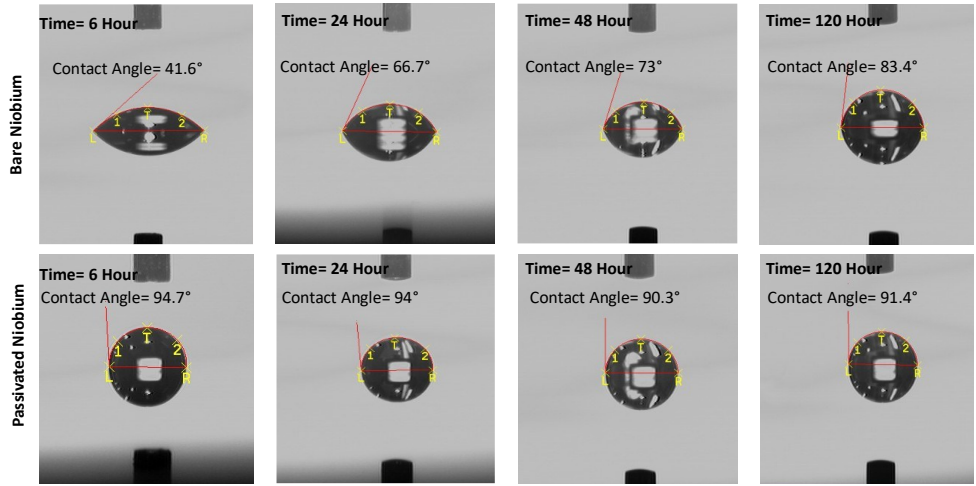

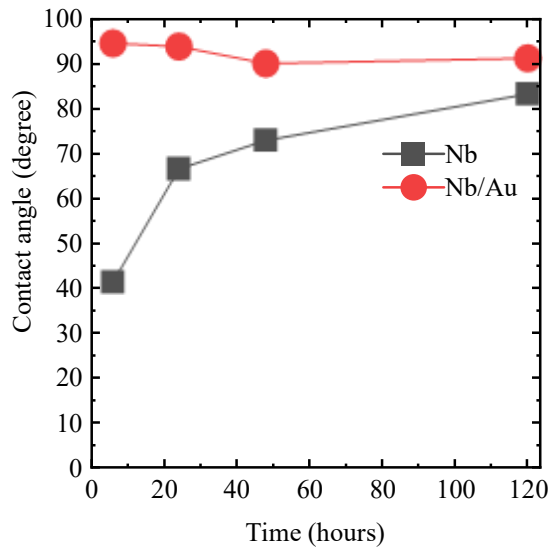

Figure S1. (Top) Photographs of contact angle measurement at different time instance for bare Nb and Au-passivated Nb. (Bottom) Time dependence of contact angles for bare Nb film and Nb films passivated with Au.